\documentclass{ieeeaccess}
\usepackage{cite}

\usepackage{booktabs}
\usepackage[official]{eurosym} 
\usepackage{float} 
\usepackage{url}
\usepackage{balance} 
\usepackage{setspace}
\usepackage{amsmath,amssymb,amsfonts}
\usepackage{algorithmic}
\usepackage{graphicx}
\usepackage{hyperref}
\usepackage{textcomp}
\usepackage{color}
\usepackage{xcolor}
\usepackage[absolute,overlay]{textpos}

\def\BibTeX{{\rm B\kern-.05em{\sc i\kern-.025em b}\kern-.08em
    T\kern-.1667em\lower.7ex\hbox{E}\kern-.125emX}}

\begin{document}
\doi{}

\title{The security implications of\\data subject rights}

\author{\uppercase{Jatinder Singh} and \uppercase{Jennifer Cobbe}}
\address{Compliant \& Accountable Systems Group, Department of Computer Science \& Technology (Computer Laboratory)\\
University of Cambridge, UK. (firstname.lastname@cst.cam.ac.uk)}

\markboth
{}
{IEEE Security \& Privacy}

\begin{abstract}

Data protection regulations give individuals rights to obtain the information that entities have on them. However, providing such information can also reveal aspects of the underlying technical infrastructure and organisational processes. This article explores the security implications this raises, and highlights the need to consider such in rights fulfillment processes.

\end{abstract}


\maketitle

\begin{textblock*}{20cm}(0cm,1cm) 
\centering{\textsf{\footnotesize{\textcolor{red}{To appear in IEEE Security \& Privacy (doi:10.1109/MSEC.2019.2914614)}}}}
\end{textblock*}
%

\section{Introduction}
\label{intro}

Recent years have seen multiple security incidents and unauthorised data disclosures. As a result, awareness of cyber security issues have been increasing, and security has become a board-level concern~\cite{cybersecboard}. 
Companies understandably wish to protect themselves against crime, attacks, and other sources of security problems. The desire to avoid the fall out of security breaches, in terms of interruption, reputational damage and financial loss is a major concern.

There are also drivers from the legal requirements around data security. In particular, European data protection law requires that those responsible for processing personal data \textit{(`data controllers')} implement data protection principles in system design and take appropriate technical and security measures to guard against the risks of data breaches. The same law also provides for various rights of individuals \textit{(`data subjects')} with regard to personal data relating to them, including the right to obtain details regarding their personal data,
and, in certain circumstances, to transfer some data to another data controller.

Previous work by others has considered the relationship between data protection by design requirements, privacy concerns, and subject access rights~\cite{clash}. This revealed a trade-off made by some data controllers in pursuit of the principle of data protection by design, often viewed shortsightedly, inadvertently or perhaps deliberately by data controllers as a confidentiality problem to be addressed by  ``Privacy Enhancing Technologies''. Taking this narrow, rather than more holistic approach to data protection
resulted in hindering data subjects from exercising their rights.

This paper considers aspects of data subject rights as they relate to security, in the context of the EU's General Data Protection Regulation (`GDPR')~\cite{gdpr}.
Specifically, we explore the security aspects as they relate to the individuals involved in a particular exercise of rights and also those that relate to controller obligations. 
The first of these concerns potential security issues resulting from the process of exercising and fulfilling these rights, an area of which there appears some awareness. 
The second relates to potential security issues resulting from the data provided in response to rights requests, and what that might reveal about the data controller's technical infrastructure and organisational processes. 
Generally, the security implications for controllers in fulfilling rights does not appear to be widely considered.

In all, we argue that in meeting data subject rights (focusing on the rights of access and portability), data controllers need 
also to consider any
security risks in fulfilling such rights; in this case, particularly as they relate to making disclosures about the controller's technical infrastructure or organisational processes. 
However, as we emphasise, this is not a trade-off:
data controllers must both fulfil data subject rights \textit{and} meet security obligations. 
Failure to do so can have consequences for controllers, not only by way of potentially significant penalties and other regulatory sanctions, but more broadly through reputational harm, loss of business, financial damage, and so forth.
By raising awareness of these issues
our aim is to encourage the development of practices and mechanisms that \textit{both} facilitate the exercise of data subject rights while also ensuring the security of data and processing activities.

\section{GDPR: Rights and obligations} 
\label{gdpr}

The General Data Protection Regulation (`GDPR') establishes a legal framework governing the processing of personal data. \textit{Personal data} is any information relating to an individual who can be identified, either directly or indirectly, from that data (or from that data in combination with other data) (GDPR Art. 4(1)). Personal data may include not only names, addresses, ID numbers, and so on, but also potentially device IDs, IP addresses (dynamic or static), online identifiers, and a range of other data relating to the specific characteristics of the individual. 
\textit{Processing} refers to any operation performed on personal data, including collection, storage, retrieval, consultation, alteration, adaptation, and use (Art. 4(2)).

Under GDPR, any natural or legal person, public authority, agency, or other body involved in processing personal data acts as a data controller or as a data processor. Data \textit{controllers} are the entities which determine the means and purposes of processing (Art. 4(7)).
Data \textit{processors} are any entity that processes personal data on behalf of and under the direction of a data controller (Art. 4(8) and Art. 28(2)), and should assist data controllers to meet their obligations (Art. 28(3)(f)).

\subsection{A basis in rights}

Under EU law, data protection is considered to be a fundamental right distinct from and of equal significance to the right to privacy \cite{deAndrade}. While GDPR came into force in 2018 and brought increased attention to the subject, data protection as a concept and as a right has a long history \cite{ustaran}. 
Human rights treaties such as the European Convention on Human Rights in 1950, while not explicitly providing for data protection, established a foundation for various countries to develop early forms of data protection frameworks in the 1950s and 60s. 
The Convention on Data Protection, agreed by the member countries of the Council of Europe in 1981, followed the OECD's Guidelines on the Protection of Privacy and Transborder Flows of Personal Data, published in 1980. The UK passed its first Data Protection Act in 1984, and the Data Protection Directive in 1995 established data protection obligations at EU level for the first time. The EU's Charter of Fundamental Rights, including data protection in Article 8, was passed in 2000 and took full legal effect in 2009 alongside the Treaty of Lisbon.

\begin{figure}[htb]
\includegraphics[width=\columnwidth]{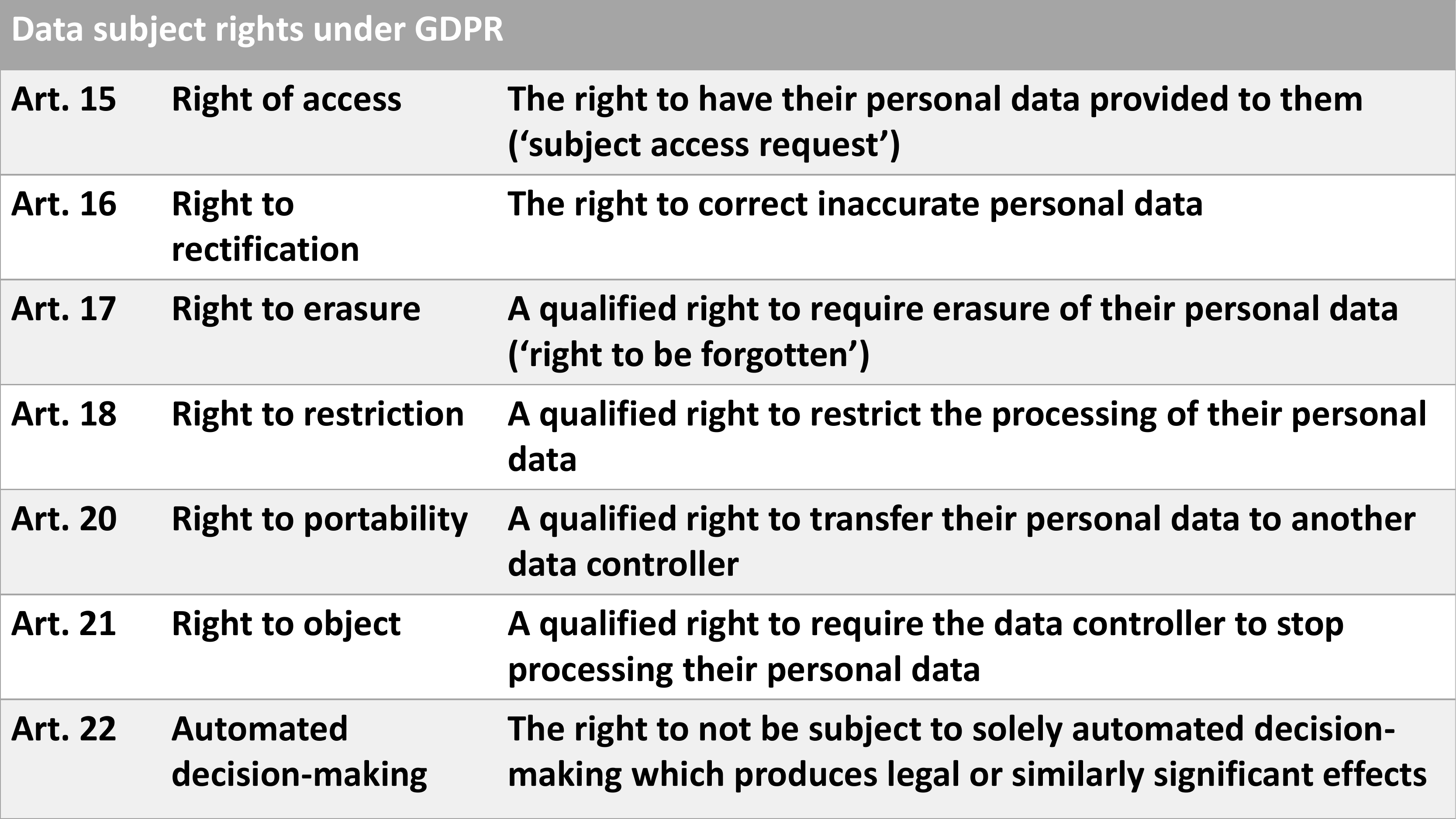}
\centering
\caption{Data subject rights provided for under GDPR. Qualified rights can be exercised when certain conditions are met (e.g. where processing is taking place under a particular specified legal basis).}
\label{gdprrights}
\end{figure}

GDPR, in expanding upon the fundamental right to data protection as established in the Charter, affords several rights to data \textit{subjects} (the individuals to whom personal data relates) (Art. 4(1)). These rights are summarised in Fig.~\ref{gdprrights}.
While other rights may have security implications (for example, where data subjects make targeted erasure requests), our focus here is on those rights which, upon their exercise, generally result in the transfer of data from data controller to data subject -- the right of access (Art. 15) and the right to portability (Art. 20).

\newlength{\xfigwd}
\setlength{\xfigwd}{\columnwidth}

The \textit{right of access} (sometimes called `subject access') provides that data subjects have a right to be told whether a data controller is processing personal data relating to them, and, where that is the case, access to that personal data and to an array of other information (Art. 15). This information includes, among other things, the purposes of processing, the categories of personal data being processed, the recipients or categories of recipient to whom the personal data has been disclosed, and the period for which the data will be stored. To fulfil this right, the data controller should provide a copy of the personal data being processed, but this should not adversely affect the rights and freedoms of others. 

The \textit{right to portability} is a more limited and qualified right than the right of access. 
It establishes that data subjects have the right to the receive in a structured, commonly used, and machine-readable format that personal data that they have provided to a data controller, or to transmit that data to another controller without hindrance from the original controller (Art. 20).
This right applies where the data is being processed either with the data subject's consent or as part of a contract between them and the data controller, and only where the processing is being carried out by automatic means.

\subsection{Obligations imposed: rights AND security}

As well as providing for these data subject rights, GDPR also imposes various obligations on data controllers and processors, both in relation to fulfilling data subject rights and in relation to security. These obligations come with potentially severe penalties for failure to comply.

Data controllers are under an obligation to facilitate data subjects in exercising their rights and to fulfil requests to exercise those rights (Art. 12). In relation to subject access 
 and portability requests, they must provide information without undue delay and in a concise, transparent, intelligible and easily accessible form (Art. 12(1)). This can be done in writing or electronically. Data controllers can only refuse to fulfil these requests if they can show that they aren't in a position to identify the data subject (Art. 11).
 
Security is also a core data protection principle (Art. 5(1)(f)). Data controllers and processors are obliged to take technical and organisational security measures appropriate  for managing the risk to the rights and freedoms of natural persons incurred in processing personal data (Art. 32). In assessing the risks incurred in processing, particular attention should be paid to the risks of accidental or unlawful 
destruction, loss, alteration, unauthorised disclosure of, or access to personal data. Technical and organisational measures could include, depending on the risks, for example, pseudonymisation and encryption of personal data; ensuring the confidentiality, integrity, and resilience of processing systems; clearly defined staff roles, policies and procedures;  and processes for routinely testing and evaluating the effectiveness of technical and organisational security measures (Art 32(1)).

GDPR's enforcement regime provides a strong incentive for controllers to manage data properly, to meet security obligations, and to fulfil data subject rights. Indeed, the possibility of fines of up to 4 per cent of global turnover for violations under GDPR (Art. 83(5)) has captured mainstream attention. But fines of that severity are likely to only be imposed for the most serious and sustained breaches, and only for certain types of violation. While up to the greater of \euro20m or 4 per cent of global turnover is indeed the maximum fine for failures to meet various obligations, including failure to fulfil data subject rights, other penalties are available to regulators. For failure to meet other obligations, including those relating to security, fines of up to the greater of \euro10m or 2 per cent of global turnover may be imposed (Art. 83(4)). Regulators also have various other enforcement powers in addition to fines, which include ordering data controllers to fulfil data subject rights, to order data controllers to bring their processing operations into compliance within a specified period, and to ban data controllers from processing personal data (Art. 58). The latter, in particular, may have severe consequences for data controllers. 

\section{Considerations in fulfilling rights requests} 
\label{exercising}
An individual exercising their rights can represent a serious undertaking. 
This means that a data controller improperly handling a request can have significant implications for the individuals involved.
For instance, access and portability requests typically reveal a sizeable quantity of sensitive information: in a portability context potentially including all relevant data ever given to the provider, and in an access context, potentially also including data which has not been given by the subject to the data controller. 
Such data often represents far more than is ordinarily accessible through  general means (e.g. use of that service).
Similar concerns are also relevant to other rights, e.g. the right of erasure might result in irrevocable damage to an individual if actioned incorrectly, and so forth.

It follows that there are security considerations for controllers regarding the handling of a rights request, including relating to the validation of the requester's identity, processing of the request itself, and the means by which the result is returned. Indeed, a controller fulfilling data subject rights would itself constitute processing, and would be subject to the same security requirements as any other form of processing.

\subsection{User authentication}
It is important for a controller to verify the identity of the individual making the request.
Means for \textit{authentication} are paramount for ensuring the requestor is who they claim, and should be employed before any request is actioned. If the data controller has reasonable doubts about the identity of a data subject, they are permitted to request further information so as to assist identification (Art. 12(6)).

Data can be processed where the identity of the data subject may not be (directly) known.
Where the controller can demonstrate that this is the case, they are not obliged to acquire or process personal data solely for the purpose of identification (Art. 11). However, they may not refuse additional data offered by the data subject to assist with identification (GDPR Recital 57).  We have previously argued that mobile platforms should provide means, including those  facilitating identification, to assist subjects in exercising their rights against app developers~\cite{claw}.

In a digital context, data subjects should normally be able to use their usual credentials to authenticate and prove their identity (Recital 57).
At a minimum, where the data subject can login to a service using their credentials, authentication for fulfilling access and portability requests should at least maintain the same level of security as for the normal user login process. In some cases, where the data involved may be particularly revealing,  it may be good practice to employ means for more stringent authentication. 
Where more than one method of contacting the data subject is available, the use of multi-factor authentication (perhaps coupled with request validation, see below) may be appropriate \cite{kennedy}. 

However, data controllers should be aware that they can only use reasonable means to identify data subjects, particularly in the context of online services (Recital 64). According to the guidance\cite{icosar} from the UK's data protection regulator, the Information Commissioner's Office (ICO), reasonableness will depend on the circumstances. Where the identity of the requester is known to the data controller, for example, they shouldn't request substantially more information (perhaps particularly where the data controller has an ongoing relationship with the requester). On the other hand, where there is some reasonable uncertainty about the identity of the requester, it would be prudent for the data controller to request identifying information; for example, where the requester uses an email address known to the data controller but requests that the personal data be sent to a postal address other than the one on file. The potential harm caused by inappropriate disclosure should also be a consideration in assessing reasonableness, which can include whether the personal data involved is particularly revealing -- health information a prime example. Note that while this guidance relates to the subject access process prior to GDPR, the reasonableness requirement is substantively the same. 

Controllers are therefore not permitted to require data subjects to submit to unreasonable identification processes, such as to provide more information than would be necessary to confirm their identity. This means that, if adopting an additional level of security for authenticating users for access and portability requests, data controllers must ensure that they do not impose a burden on data subjects seeking to exercise their rights, or request excessive identifying information. 

\subsection{Processing the request}
Related to identity, it would be good practice in some contexts to further validate that it was in fact the subject that made the request~\cite{claw}. For instance, in an online context, emailing a confirmation link to the account's registered email address can help avoid situations where someone's account with the service (but not their email) has been compromised, or where a login-session has been hijacked after failing to logout of a shared machine, their phone has been left unlocked, and so forth.

Once the requestor's identity and request has been validated, the next step is to ensure that the request is responded to appropriately. 
While controllers are obliged to fulfil requests, they are permitted to refuse requests where they are manifestly unfounded or excessive (Art. 12(5)). This provides a means, for example, to prevent something of a `denial of service' by way of repetitive requests.
Should controllers refuse to fulfil a request on these grounds, they are obliged to give reasons demonstrating that this is the case (Art. 12(5) and Recital 59). 
Note that data subjects are not and should not be obliged to state their motives in making a request. In practice, controllers may find other factors indicative, such as the character and timing of the requests, particularly those frequently repeated. 

Naturally, it is important for controllers to have processes in place to ensure the proper handing of the request, that the data requested relates to the data subject, and that, through quality assurance, data is validated before it is returned.
While these seem exceedingly obvious, data breaches have resulted from improperly handling rights requests -- in one reported example, the results of an access request included the data pertaining to a different individual that had also made a request~\cite{tristanclaw}.

\subsection{Protecting the response}
The mechanism by which responses are delivered also warrants attention. 
As discussed, the results of a right of access or portability request are potentially revealing, not only because the data they include is inherently personal, but also because it represents an aggregation of personal data. 
Having access to all of one's interactions on a social media platform over a period of time, for instance, provides information beyond that accessible through general use of the account, by revealing information of hidden (`deleted') messages, blocked contacts, and so on.

As such, measures by which responses are protected, or which reduce the risk of data leakage, warrant consideration by controllers. 
Simply emailing a response (unencrypted) to the user's registered email address might be convenient and acceptable in some circumstances, but is arguably  insufficient in others -- especially if data relates to financial matters, special categories of personal data, and so on.

Less common are means to protect the data once it moves `off-platform'. This is not an explicit obligation for the controller, but  seems useful from a security point of view. One example might be to deliver the archive in an encrypted form with a user-specified password, so that even if the archive is leaked its contents  still have some degree of protection. 

\subsection{Implementation}

It appears that some organisations have given consideration to such issues. 
Our experience in exercising our rights with various organisations has provided evidence of some common measures being taken, including:
\begin{enumerate}
    \item Ensuring the request and download of information occurs through the controller's platform, meaning users are validated through the platform's standard access control procedures (although requests made via another method, for example email, must still be fulfilled).
    \item Sending an email to the account's registered address notifying that a request has been made, and sometimes requiring that user to confirm the request, as a means for alerting data subjects and mitigating any issues should their account (and thus the request, though not their email) be compromised.\footnote{A confirmation step seems more appropriate than simply alerting, as it gives some extra assurance that the data subject actually made the request, while discouraging opportunism (e.g. someone leaving a machine `logged-in'). On one social media platform we observed that an email was sent immediately on request, notifying that the request was made through their platform, and stating that the account may be compromised if this request was unfamiliar.  The time taken from receiving the request notification until the entire information archive became available was \textasciitilde19min, and the archive was directly accessible through the platform (no email access required). This gives very little time for a user,  whose account may be compromised, to respond.}
    \item Having the information archive only downloadable for a short time after the request, to restrict the possibility for distribution or future leakage.
\end{enumerate}

\section{Organisational security concerns} 
\label{implications}

The security considerations just described concern the protection of personal data related to making and fulfilling a request.
In these cases, where a failure occurs, the data breach (or other issue) typically concerns those data subjects to whom the request (or response) relates.
There is some awareness of these problems -- as shown by guidance on related issues produced by Supervisory Authorities (data protection regulators) around, for example, identity verification~\cite{icosar},
the GDPR recitals, and so forth.

Less discussed in a rights context, however, are the potential \textit{security risks facing the controller}.
The data returned as part of an access or portability request might reveal aspects of the controller's technical infrastructure and its implementation, and could also  indicate the nature of the organisational processes that are in place.
Revealing such information can pose a security risk, as it may help in facilitating a cyber attack.

This bears consideration, not least because a security incident affecting the controller can be systemic and result in harms \textit{at scale}. 
This has direct data protection implications -- for example, a security breach might leak data on \textit{all} data subjects (as opposed to just those relating to a request (per \S\ref{exercising})). Note also that security issues extend beyond data protection,  e.g. where issues of system downtime can impact users, possibly with significant consequences, such as where the service represents some critical infrastructure.

\setlength{\xfigwd}{\columnwidth}
\begin{figure*}[tb]
\includegraphics[width=50em]{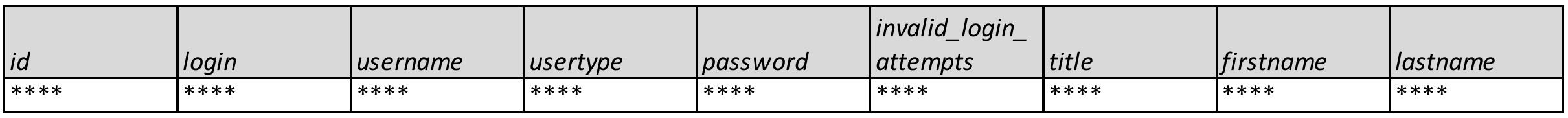}
\centering
\caption{A subset of (masked) customer information extracted from a database, in which aspects of the database schema are visible. 
A web search based on this subset of field names makes clear this represents the use of a popular e-commerce software package.
This information could then be used to facilitate an attack, for instance, through a known vulnerability in that software package, by aiding social engineering, etc.}
\label{dump}
\end{figure*}

\subsection{Cyber reconnaissance}
Information gathering is an important stage of a cyber attack. 
That is, knowledge about a potential target---be it of their technical infrastructure or organisational processes---helps facilitate an attack~\cite{hackgen}.
Having information about the technology employed (for instance, the software used and version numbers, network configuration, open ports, and so forth) might reveal attack vectors, exploitable systems to target once the organisation is infiltrated, and potentially possible countermeasures or protections employed.  Similarly, knowledge of the social and organisational practices of an organisation can guide strategy around the means for the attack~\cite{enisa}, expected response time, investigative capability, vulnerable business processes and susceptible staff, and so forth. 

At a technical level, it is common for networks to be used for nefarious information gathering purposes, since a network renders a remote system reachable.
Major technology service providers report thousands of cyber incidents each day, often detecting  the use of automated tools that scan and probe both for information about the host being queried and to indicate potential vulnerabilities (for example, outdated software, improper configuration, and so on). Note that such techniques are used defensively (e.g. by organisations seeking to improve their security) and offensively (e.g. by cyber criminals).
Towards this, various intrusion detection and prevention tools involve monitoring to uncover and mitigate imminent threats (including reconnaissance efforts), or actual security violations (see~\cite{nist}).
Such operations might be targeted at a specific organisation, or could be more `scatter-gun', e.g. where a whole series of network addresses  could be probed to find potential targets. 

However, not all information gathering methods are technical~\cite{enisa}. Cyber crime operations are known to use people as part of their attack strategy. This could involve  social engineering (i.e. deceiving a staff member to reveal particular information or take an action)~\cite{socialeng},  a malicious insider, which may be bribed, extorted, or perhaps a `plant' employed specifically to facilitate an attack~\cite{alice}.

\subsection{Exposing the organisation}
In this rights context, an interesting consideration is that a controller, by providing information, could potentially expose the underlying technology and processes of the organisation. 
In other words, there is the potential that the information provided by fulfilling access and portability requests could assist cyber reconnaissance. 
Indeed, these data access rights present a new method for `information gathering', providing a legal mechanism whereby potentially deep organisational internals are exposed -- information that would otherwise be generally inaccessible externally or difficult to obtain.

Of course, any risk depends on circumstances, and in particular, the nature of the request and information returned in the controller's response.
That said, there appears to be little discussion of such risks, meaning that organisations may well be, inadvertently, leaking information that might have security implications. 
The fact that controllers are providing direct data exports from technical components~\cite{mirrors,tristanclaw} suggests these aspects have had little consideration.

\subsubsection*{Illustrating some risks}
We now illustrate some of the possible risks that might result from the responses to access and portability requests. 
Our discussion here is only indicative; again, any risks will depend on the particular circumstances.

First,  the response can indicate the software, services, and technologies used, particularly where the data forms an `extract' directly from a technical system.
Such details might be revealed directly or indirectly, e.g. through metadata included in a report describing the package or version number, as a screenshot where the interface indicates the software used, or perhaps where certain data structures and formats imply the use of a particular products or services (see Fig~\ref{dump}).
Such information could assist technical attacks, where knowledge of the software stack and its vulnerabilities might facilitate an exploit.
It can also provide the background for facilitating spearphishing attacks (attacks targeting particular staff members, e.g. through fake emails about or resembling those common to the application used), social engineering (e.g. by masquerading as a technical support with detailed knowledge of the software used), and other forms of deception (see~\cite{socialeng}).

Another consideration is whether the data returned can indicate some of the risk mitigation measures that the controller employs. 
For instance, if logging\slash audit records are revealed (as they often should be if they are associated with a data subject), it can help indicate some of the organisation's security strategy: Are failed-logins recorded? Do such records include the IP address? Is every input made throughout their web application recorded? 
These aspects can influence a would-be attacker's strategy; for instance, if few records appear to be kept, a bolder systems-interrogation approach might be encouraged as it suggests a lack of capability by the organisation to detect and investigate such.

It is reportedly common that a direct extract (`dump') from the database is returned as with a request~\cite{mirrors}.
In addition to including the personal data, generally such information  will also reveal the database \textit{schema}---the way information is structured---including the fields of data recorded and their names, while the data itself can suggest the fields' data types.  
The example presented in Fig.~\ref{dump} shows how a schema reveals the software used, which, as discussed earlier, can have implications.

The schema might also reveal some of the security measures employed. 
As above, database tables or fields that appear dedicated to logging or audit might indicate the aspects of an application being monitored (and the aspects that aren't!). 
Regarding more organisational measures, the structure could suggest, for instance, over what aspects staff give oversight --  an \texttt{[Approved\_by]} field in a table dedicated to transactions might imply manual staff verification, or if a table has separate fields regarding security verification answers (a poor design), e.g. \texttt{[Maiden\_Name]},  \texttt{[Fav\_Song]}, \ldots, revealing this indicates the background information required to impersonate a particular target. 

From a technical perspective, knowledge of the schema, while not an attack vector itself, can provide the background to assist in enabling system compromises.
One common attack type is \textit{injection}~\cite{enisa},
 which essentially involves running malicious queries against a datastore. 
 These queries can entail reads (information leaks) or writes (modification and deletion). Although injection attacks are well-known, and mitigated by  best-practices, they still represent an extremely common vulnerability -- ENISA recognising it as the top form of web application attack~\cite{enisa}. As datastore queries execute in line with the datastore's schema, knowledge of the schema can enable attacks (be they through injection, or some other vector) to be more focused and streamlined through predefined and targeted queries.

The above represent but a few examples of some potential security implications of rights requests that reveal the controller's underlying systems infrastructure.
Again, any risks depend on the specific circumstances; including the type of request, data actually returned, the technical stack, organisational processes, and so forth.

\label{exposing}

\section{Discussion} 
\label{discussion}

GDPR's data subject rights
reflect the human-rights foundations of EU data protection law. 
They exist to inform and empower people in relation to their personal data. 
As such, fulfilling rights, including the right of access and right to portability is an extremely important aspect of GDPR compliance.
 Generally, organisational security concerns  do not and cannot trump data subject access rights, and must not pose a barrier to exercising rights. 
  This means that data controllers should consider how access and portability rights requests can be fulfilled in a way that is fully compliant with the GDPR's requirements, but that also minimises the associated security risks. 
We now consider some potential ways forward.

\subsection{Raising awareness}
Data controllers, as part of their security obligations, should consider the security aspects of fulfilling data subject rights. 
This cannot mean using security as a reason to not fully comply with requests.
However, it should mean that data controllers are aware of and take steps to mitigate the potential security risks which could arise in doing so. 
In other words, controllers need to adopt a security-oriented mindset in complying with access and portability requests.

To date, there appears to be little awareness of  the security implications for the controller in fulfilling rights requests. 
Not only is there scant discussion of such issues (cf. \S\ref{exercising}), but, from work analysing rights responses, it also seems that many controllers are reactive in their approach to fulfilling rights requests.
For instance, providing data `dumps' from technical systems (\cite{tristanclaw,mirrors}) suggests that rigorous assessments of the risks to an organisation of fulfilling a rights request (in a particular way) were not undertaken.

It may well be that many controllers, perhaps driven by a rush to comply, were fairly ad-hoc in their responses; they may not yet have defined rights fulfilment processes, let alone subjected them to a rigorous security analysis. 
It's also conceivable that the rights requests were fulfilled by a team or business unit that does not ordinarily conduct security risk assessments, and did not consult with those undertaking such assessments. One may expect such issues to occur more with smaller firms, as larger firms may be more likely to have in-house security expertise, measures for evaluating new business operational processes, and so on; though larger firms might have greater fragmentation amongst business groups, giving more scope for things to `slip through the cracks'.

Regardless of the reasons, as we have outlined, there can be risks in the way rights requests are served. As such, there is a real need to raise awareness of these issues, so that controllers can engineer their rights fulfilment processes to account for possible security concerns. Indeed, this is a key aim of this article.

One starting point might be through information provided by Supervisory Authorities. As mentioned, these authorities already issue guidance on fulfilling rights requests~\cite{icosar}. Having guidance documents which also indicate that fulfilling subject rights requests can bring security risks could alert and encourage those dealing with requests to consider such issues and seek advice and expertise where necessary.

\subsection{Mode of presentation}

GDPR provides for rights of access and portability in relation to eligible personal data which is being processed. However, in fulfilling these requests, the \textit{data provided to data subjects need not necessarily reflect the underlying technical infrastructure}, so long as all of the required data is in fact provided. At its simplest level, controllers do not need to provide the actual table headers, database schema, software metadata (name, version number), and other potentially revealing information.
In some cases, security concerns could mean that obfuscating the underlying technical infrastructure in some way is a necessary measure to help data controllers continue to meet their security-related obligations under GDPR.

The need to consider the security implications of providing responses to rights requests could also result in making the responses more useful. 
This could potentially incentivise data controllers to find ways to present the data provided in response to subject access requests in such a way as to be more understandable and intuitive for data subjects. At a simplest level, this could entail translating column headers from a technical specification to names that are more descriptive. 
More creative approaches are certainly possible. However, a concern is that data controllers might use the presentation of information in an attempt to minimise or reduce the prominence of data they consider inconvenient. It should be emphasised that, regardless of the format chosen for displaying data, all data should be returned and controllers \textit{must not mislead} data subjects about what data is being processed.

Portability necessarily involves representing the data in a technical format. 
GDPR's requirements that data provided in response to portability requests is given in a structured, commonly used format gives fewer possibilities for curation.
However,  even a technical representation need not include extraneous system-related data (e.g. software version numbers), nor use the same descriptors as the actual underlying system; and indeed, requirements for data portability to be in a common format could well make such extraneous or specific information inappropriate. Further, there may well be scope for making the data more `technically' useful rather than strictly representing that of the underlying technical system.

\subsection{Standards and practices}
The development of standards, and `best-practices', as they relate to serving requests can assist as (i) they not only encourage firms to consider and develop processes for fulfilling data requests, but (ii) also likely entail some translation (and thought) in mapping the local data, systems and processes to that of the standard.
This can help either explicitly, where implementing a standard provides an opportunity for the organisation to consider the security risks as it defines its policy, and also implicitly, where adhering to the standard might naturally work to obfuscate (some) the underlying details, including removing the inclusion of extraneous metadata that could otherwise be obtained through `technical dumps'. Ideally, security considerations should feature as part of any rights-oriented standards development processes.

\subsubsection{Interoperability standards}
Though standards, and the translation they entail, can generally assist such issues, they are especially relevant for the right of the portability.
The motivation for the right is to encourage competition, as a means to enable data subjects to migrate to other services (controllers) should they wish. The Art. 20 right to portability requires (i) controllers to provide data in a technical format, and (ii) where technically possible, to enable a controller to directly transfer the subject's data to another service. Interoperability standards are already being discussed, as a means to facilitate both these aspects. Indeed, GDPR suggests that data controllers should be encouraged to develop interoperability standards (Recital 68), and Supervisory Authorities may have a role to play. 

Platform-based approaches to common standards are beginning to emerge. For example, there is the Data Transfer Project (see \url{https://datatransferproject.dev/}), which is an open-source project primarily led by the tech giants.
Its key aim is to facilitate the `direct transfer' aspect of the right to portability between service providers, and security is mentioned as a key consideration. 
Though a direct transfer regime brings efficiency and can preclude disclosing the raw data back to data subjects (if the subjects desire direct transfer), subjects still maintain the right under portability to have access to this information.  
As portability platforms develop, facilitating transfer necessarily requires standards and processes to be implemented -- which again provide opportunities for organisations to consider and reduce their security risks.

\subsubsection{Emerging best practice}
There are also projects underway that aim at providing platforms to assist data subjects in exercising their rights, by giving information and tooling.\footnote{For an indicative list, see \url{https://datarights.wiki/index.php/List_of_all_access_request_projects}.} In line with the awareness aspect already discussed, these subject-oriented (i.e. people-focused) platforms also bring visibility over the practices of various organisations, by operating across them. In turn, this can encourage better practices (de-facto standards), though coordinating subject demands, and where controllers can learn from others as to the appropriate (and inappropriate) information to include in requests.  

More generally, organisations do not operate in a vacuum -- knowledge and approaches are often shared. 
Therefore it may also be beneficial for data controllers to exchange and share knowledge of practices in relation to data subject rights and security, perhaps through professional or industry bodies, as well as through more informal forums. 
This would allow experience and knowledge to transfer between organisations, ideally raising the standards of industry as a whole.

\section{Concluding remarks}
Data rights are a fundamental aspect of data protection regulations.
And while organisations increasingly recognise the importance of cyber security, 
 so far there has been little discussion regarding the security implications of data subject rights.
Here we have argued that controllers should consider and account for the possible security risks as they fulfil rights requests. 
It has been observed that there are tensions between privacy mechanisms and exercising rights~\cite{clash,claw}, where confidentiality mechanisms can inhibit rights fulfilment.
What we have discussed is in a similar space; though we focus on where fulfilling right requests has resulting security implications. 
To reiterate, this should not be considered a trade-off between rights and security, not only because of the importance of rights, but also because many risks can be mitigated through integrating some security-thinking into the methods and processes by which rights are fulfilled. 
Rather, awareness seems the key concern -- to date, there appears little discussion of such issues. 
And while improvements may occur naturally as organisational approaches to data protection mature,
at present, rights-related security issues warrant greater attention.

In all, this appears to be an opportunity. A more security-oriented approach to rights fulfilment works to improve controller security, which in turn can help prevent attacks, hacks, and possible breaches that can have wide-reaching consequences for data subjects, controllers and beyond.
At the same time, forcing more attention on the way information is delivered to data subjects also provides an opening to consider how such responses can be more useful and meaningful.

\section*{Acknowledgements}
We thank Tristan Henderson, Jef Ausloos and Rejo Zenger for their results regarding controller request fulfillment.
We acknowledge the financial support of the UK Engineering \& Physical Sciences Research Council (EPSRC) grant EP/P024394/1, and the University of Cambridge, via the Trust \& Technology Initiative.

\bibliographystyle{plain}

\begin{IEEEbiography}  {Jatinder Singh} is based at the Department of Computer Science and Technology (Computer Laboratory), University of Cambridge, where he leads the Compliant and Accountable Systems Research Group. The group focuses on the intersections of CS and law, which from a technical perspective spans areas including security, privacy, data management, and auditing, typically in the context of cloud and distributed systems (IoT). He is also a Fellow of the Alan Turing Institute, the UK's national centre for data science and artificial intelligence, and co-chair's Cambridge's Trust \& Technology Initiative. He received his PhD in Computer Science from the University of Cambridge and has some background in Law.
\end{IEEEbiography}

\begin{IEEEbiography} {Jennifer Cobbe} is a researcher in the Compliant and Accountable Systems Group in the Department of Computer Science \& Technology at the University of Cambridge, and the coordinator of Cambridge's Trust \& Technology Initiative. She holds a PhD in Law and an LLM in Law and Governance from Queen's University, Belfast; for her PhD, she studied machine learning in commercial and state internet surveillance, data protection, and privacy.
\end{IEEEbiography}

\EOD
\end{document}